\documentstyle[prl,aps,epsf]{revtex}
\begin{document} 
\draft
%--------------------------------------------------------------------
\title{$h$/2$e$ Oscillations and Quantum Chaos in Ballistic Aharonov-Bohm 
Billiards}
%--------------------------------------------------------------------
\author{Shiro Kawabata\footnote{
Present address: Physical Science Division, Electrotechnical Laboratory, Umezono 1-1-4, Tsukuba, 
Ibaraki 305-8568, Japan \\E-mail:shiro@etl.go.jp
} 
and Katsuhiro Nakamura}
\address{
Department of Applied Physics, Osaka City University, 
Sumiyoshi-ku, Osaka 558, Japan}
%
%--------------------------------------------------------------------
%\date{\today}
\maketitle
%--------------------------------------------------------------------
\begin{abstract} 
We study the quantum interference effect for the single ballistic Aharonov-Bohm billiard
in the presence of a weak magnetic field $B$. 
The diagonal part of the wave-number averaged reflection coefficient \( \delta {\cal R}_D \) is calculated 
by use of semi-classical scattering theory. 
In addition to the appearance of "$h/2e$ oscillation"
that are caused 
by interference between time-reversed coherent backscattering classical trajectories,
$B$ in the conducting region leads to negative magnetoresistance and dampening of the $h/2e$ oscillation amplitude.
The $B$ dependence of the results reflects the underlying classical (chaotic and regular) dynamics. 
\end{abstract}
%--------------------------------------------------------------------
\pacs{PACS numbers: 05.45.+b, 03.65.Sq, 73.20.My, 73.20.Fz}
%
%\narrowtext
%
%
%
\section{Introduction}
Recently, the interplay of chaos and quantum 
interference in $ballistic$ quantum dots has intrigued experimentalists and theorists 
alike.
The quantum interference effects, $e.g$., ballistic weak localization (BWL) ~\cite{rf:Baranger} and
ballistic 
conductance fluctuation ~\cite{rf:Jalabert} in  
such structures depend on whether the underlying classical
 dynamics is regular or chaotic. Therefore, these effects are interesting from the viewpoint 
 of the theory of  $quantum$ $chaos$ ~\cite{rf:Quantum chaos}.

 More recently, we predicted that $h/2e$ 
 oscillation of magnetoconductance, analogous to Altshuler-Aronov-Spivak (AAS) effect 
~\cite{rf:Altshuler} in disordered systems ~\cite{rf:AAS}, should be observable 
 in a single $ballistic$ Aharonov-Bohm ring (hereafter called AB billiard) with 
 magnetic flux penetrating 
 $only$ through the hollow ~\cite{rf:Kawabata}. 
 This phenomenon of conductance oscillation is caused by the interference between a pair of 
time-reversed coherent back-scattering classical paths that wind the center obstacle 
in the billiard.
We calculated the diagonal part of the BWL correction to the wave-number averaged reflection coefficient
by use of semiclassical scattering (SCS) theory ~\cite{rf:Baranger,rf:Jalabert,rf:BJS}. 
Our analysis ~\cite{rf:Kawabata} yielded for the $chaotic$ AB billiard
\begin{equation}
  \delta {\cal R}_D (\Phi) \sim \sum_{n=1}^{\infty} 
         \exp{\left( - \delta n \right)} 
	     \cos{\left( 4 \pi n \frac {\Phi} {\Phi_0} \right)}, 
  \label{eqn:aas1}
\end{equation}
where \( \Phi_0 = h/e\) is the magnetic flux quantum and \( \Phi \) is the magnetic flux 
that penetrates the hollow.
 In eq. (\ref{eqn:aas1}) \( \delta = \sqrt { 2 T_0 \gamma / \alpha } \), where $\alpha$, $T_0$, and $\gamma$ are system-dependent
 constants and correspond to the 
variance of the winding number distribution ~\cite{rf:Berry}, the dwelling time for the shortest classical
 orbit and the escape rate ~\cite{rf:BJS,rf:LDJ}, respectively. In this case, the oscillation amplitude decays 
 exponentially with increasing rank of higher harmonics $n$, so that the main contribution
 to the conductance oscillation 
 comes from the $n$=1 component that oscillates with the period of $h$/2$e$.
By contrast, for $regular$ and $mixed$ (Kolmogorov-Arnold-Moser tori and chaotic sea) AB billiard,
we obtained ~\cite{rf:Kawabata}
\begin{equation}
  \delta {\cal R}_D (\Phi) \sim \sum_{n=1}^{\infty} 
         F \left(  z - \frac1 {2} , z +\frac1 {2} ; - \frac{n^2} {2 \alpha} \right)  
	     \cos{\left( 4 \pi n \frac {\Phi} {\Phi_0} \right)}, 
  \label{eqn:aas2}
\end{equation}
where $F$ and $\beta$ are the hypergeometric function of confluent type and the exponent of 
dwelling time distribution \( N(T)  \sim T^{- z} \) ~\cite{rf:BJS,rf:LDJ}, respectively.
In eq. (\ref{eqn:aas2}) the oscillation amplitude decays algebraically for large $n$, and 
therefore the higher-harmonics components give noticeable contribution to 
magnetoconductance oscillations. 
These discoveries indicate that {\em the $h/2e$ AAS oscillation occurs in both ballistic and diffusive 
systems forming the AB geometry and the behavior of higher harmonics components 
reflects a difference between chaotic and non-chaotic classical dynamics}.

 In real experiments, however, the magnetic field would be applied 
 to the $entire$ region (both the hollow and annulus) in the billiard.
Thus it is indispensable to apply the SCS theory to this case, in order to see the comparison
between the experimental data and theoretical prediction.
In this situation, we shall envisage $h/2e$ oscillation together with {\em the negative magnetoresistance} and 
{\em dampening of the $h/2e$ oscillation amplitude} with increasing magnetic field.
In this paper, we shall focus our attention on two-dimensional ballistic AB billiards (e.g., the insets in Fig. 1) with 
the magnetic flux penetrating through the entire region and calculate reflection amplitude by use of SCS theory.
\section{Semiclassical theory}
 Following Baranger, Jalabert and Stone's arguments ~\cite{rf:Baranger,rf:BJS}, we start with a quantum-mechanical reflection 
amplitude ~\cite{rf:FL}
\begin{equation}
    r_{n,m}  =  \delta_{n,m} - i \hbar \sqrt{\upsilon_n \upsilon_m} 
    \int dy \int dy' \psi_n^*(y') \psi_m(y) 
  G(y',y,E_F), 
  \label{eqn:a1}
\end{equation}
where \(\upsilon_m(\upsilon_n)\) and \(\psi_m(\psi_n)\) are the longitudinal velocity 
and 
transverse wave function for the mode $m(n)$, respectively. $G$ is the retarded Green's 
function.
To approximate $r_{n,m}$ we replace $G$ by its semiclassical 
Feynman path-integral expression ~\cite{rf:Gutzwiller},
\begin{equation}
  G^{sc}(y',y,E) = \frac {2 \pi} {(2 \pi i \hbar)^{3/2}} \sum_{s(y,y')} 
  \sqrt{D_s} 
  \exp \left[ 
                            \frac i {\hbar} S_s (y',y,E) - i \frac \pi {2} \mu_s
                       \right],
  \label{eqn:a2}
\end{equation}
where $S_s$ is the action integral along classical path $s$, 
 \( D_s = ( \upsilon_F \cos{\theta'})^{-1} \left| ( \partial \theta /\partial y'  )_y \right| \)
, \( \theta \) (\( \theta' \)) is the incoming (outgoing) angle, and \( \mu_s \) is the Maslov index.
Assuming hard walls in the leads, we substitute eq. (\ref{eqn:a2}) into 
eq. (\ref{eqn:a1}) and carry out the double integrals by a stationary-phase approximation.
Thus we obtain
\begin{equation}
  r_{n,m}   =   - \frac {\sqrt{2 \pi i \hbar}} {2 W} \sum_{s(\bar n,\bar m)} 
  {\rm sgn} (\bar n) {\rm sgn} (\bar m) \sqrt{\tilde D_s}  
      \exp{ 
        \left[
              \frac i {\hbar} \tilde S_s (\bar n,\bar m;E)-i \frac \pi {2} \tilde \mu_s
        \right]
	  },
  \label{eqn:a3}
\end{equation}
where $W$ is the width of the hard-wall leads and \( \bar m = \pm m \). The summation 
is over trajectories between the cross sections at $x$ and $x'$ with angle
\( \sin{\theta} =  \bar{n} \pi / k W  \). In eq. (\ref{eqn:a3}), 
\( 
  \tilde{S_s} (\bar{n},\bar{m};E) = S_s(y'_0,y_0;E)+ \hbar \pi ( \bar{m} y_0 - \bar{n} y'_0 ) / W 
\), 
\( 
  \tilde{D_s} = ( m_e \upsilon_F \cos{\theta'})^{-1} \left| ( \partial y /\partial \theta' )_{\theta} \right|
\) 
and
\( 
\tilde{\mu_s} = \mu_s + u \left( -( \partial \theta / \partial y )_y' \right)
                      + u \left( -( \partial \theta' / \partial y' )_{\theta} \right),
\)
respectively, where $u$ is the Heaviside step function.
The Kronecker delta term in eq. (\ref{eqn:a1}) is exactly canceled by the contributions of paths 
of zero length ~\cite{rf:Lin}. 
Within the diagonal approximation ~\cite{rf:Baranger,rf:BJS}, the quantum correction \( \delta R \) to the classical reflection 
probability \(R_{cl}\), $viz.$,
\begin{equation}
  R = \sum_{n,m=1}^{N_M} \left| r_{n,m} \right| ^{2} \approx R_{cl} + \delta R 
  \label{eqn:refrection}
\end{equation}
with the mode number $N_{M}$, is given by
\begin{equation}
  \delta R_D   =   \frac 1 {2} \frac{\pi} {k W}
			    \sum_n \sum_{s \ne u}
				\sqrt{\tilde A_s \tilde A_u}   
	   \exp \left[
			      i k \left(
			          \tilde L_s - \tilde L_u
			           \right)
			         +i \pi \nu_{s,u}
			         \right],
  \label{eqn:a7}
\end{equation}
where $s$ and $u$ label the classical trajectories.
In eq. (\ref{eqn:a7}),
\( \tilde L_s = \tilde S_s / k \hbar \)
,
\( \nu_{s,u} = \left( \tilde \mu_u - \tilde \mu_s \right) / 2 \)
, and
\( \tilde A_s = \left( \hbar k / W \right) \tilde D_s \)
.
The wave-number averaging of \( \delta R_D \) over all $k$, denoted as
\( \delta {\cal R}_D \)
, eliminates all paths except those that satisfy
\( \tilde L_s = \tilde L_u  \)
in eq. (\ref{eqn:a7}). In the absence of spatial symmetry,
\( \tilde L_s = \tilde L_u  \)
 holds if $u$ is the time reversal of $s$. 
 A weak magnetic field does not change the classical trajectories appreciably but does change the
 phase difference between the time-reversed trajectories by \( (S_s - S_u)/\hbar = 2 \Theta_s B / \Phi_0 \), where \( \Theta_s \equiv 2 \pi \int_s {\bf A} \cdot d{\bf \ell}/ B \) is the effective area almost 
 enclosed by the classical path. 
 To evaluate the summation over $s$ and $n$, we shall reorder the 
 backscattering classical paths according to the increasing effective area. 
 Therefore,
 we obtain
\begin{equation}
  \delta {\cal R}_D (B) \sim \int_{-\infty}^{\infty} d \Theta N(\Theta) 
       \exp{ \left( i \frac {2 \Theta B} {\Phi_0} \right) },
  \label{eqn:a9}
\end{equation}
where \( N(\Theta) \) is the distribution of $\Theta$. 
The phenomelogical statistical theory leading to a distribution of the enclosed 
area $N(\Theta)$ for chaotic AB billiards
is given as follows.
There exist two kind of classical paths 
and $N(\Theta)$ is the sum of the distribution of unwinding trajectories, viz.,
$N_0(\Theta)$ and that of $n (\ne 0)$ winding trajectories.
This is due to all classical trajectories winding
a center obstacle $n$ times until they exit, except for very short backscattered paths ($n$=0 component).
$N_0(\Theta)$ is essentially the same as that of  
an ordinary chaotic billiard ($e.g.,$ stadium), obeying a monotonic 
exponential law ~\cite{rf:BJS,rf:LDJ}, $i.e$., \( N_0(\Theta) \sim \exp ( -\varepsilon_{cl} \left| \Theta \right|) \)
, where \( \varepsilon_{cl} \) is the inverse of the average area enclosed by 
classical trajectories.
Therefore, the full distribution of the enclosed area is given by
\begin{equation}
    N(\Theta)  \sim  
	             N_0 (\Theta)
	           + 
				 \sum_{\stackrel{n=-\infty}{n \ne 0}}^{\infty}  
				N(\Theta,n) P(n),
  \label{eqn:b1}
\end{equation}
where \( P(n) \) and \( N(\Theta,n) \) are the distribution of the winding number \( n \)
and that of the enclosed area for a given \( n \), respectively.
Owing to the ergodic properties of fully chaotic systems, \( N(\Theta,n) \) 
is assumed to obey a Gaussian distribution in which the variance of area is proportional to \( n \)
, $i.e$., 
\begin{equation}
N(\Theta,n)	\sim \frac{1} {\sqrt{2 \pi \beta \left| n \right|}}
					 \exp{
	                   \left[
				              - \frac{{\left( \Theta - n \Theta_0 \right)}^2} 
							          {2 \beta \left| n \right|}
	                   \right]. 
					   }
  \label{eqn:b2}
\end{equation}
On the other hand, exploiting Berry and Keating's argument ~\cite{rf:Berry}, \( P(n) \) is given by
\begin{equation}
    P(n)  =  \int_0^\infty dT P(n,T) N(T)
	      \sim \exp { \left( - \delta \left| n \right| \right) },
  \label{eqn:b3}
\end{equation}
where $\delta=\sqrt{ 2 T_0 \gamma / \alpha}$.
In eq.(\ref{eqn:b3}),
$N(T) \sim \exp (- \gamma T)$
and 
\begin{equation}
P(n,T)  = \sqrt{ \frac{T_0} {2 \pi \alpha T}} 
     \exp \left( -\frac{n^2 T_0 } { 2 \alpha T} \right)
  \label{eqn:b4}
\end{equation}
are the exponential dwelling time distribution ~\cite{rf:BJS,rf:LDJ} and the  
Gaussian distribution of winding numbers $n$ for trajectories with a 
fixed \( T \) ~\cite{rf:Kawabata,rf:Berry}, respectively. 
With the use of eqs.(\ref{eqn:b2}) and (\ref{eqn:b3}), we reach 
\begin{equation}
    N(\Theta)  \sim  A e^{-\varepsilon_{cl} \left| \Theta \right|}
	           + 
				\sum_{\stackrel{n=-\infty}{n \ne 0}}^{\infty} 
					 \frac 1 {\sqrt{2 \pi \beta \left| n \right|}}
					 \exp{
	                   \left[
				              - \frac {{\left( \Theta - n \Theta_0 \right)}^2} 
							          {2 \beta \left| n \right|}
	                          - \delta \left| n \right|
	                   \right]
					   }.
  \label{eqn:a10}
\end{equation}
To examine the validity of expression (\ref{eqn:a10}), 
we directly calculated $N(\Theta)$ for chaotic AB 
billiards (a single Sinai billiard ~\cite{rf:Sinai})
by mean of classical numerical simulations.
In the calculations, we inject \( 10^8 \) particles into the billiard 
at different initial conditions.
\( N(\Theta) \) for the chaotic AB billiard has proved to be nicely 
fitted by eq. (\ref{eqn:a10}) [see Fig. 1(a)].
(In this case, \( \Theta_0 / 2 \pi \) is approximately the average area between the outer square and 
inner circles.)
As the size of center obstacle approaches zero, the Sinai geometry becomes square (regular billiard).
In this case, we have confirmed that the oscillation structure disappears and \( N(\Theta) \) obeys a well-known
power law ~\cite{rf:BJS,rf:LDJ}.
Substituting eq. (\ref{eqn:a10}) into eq. (\ref{eqn:a9}), we finally obtain 
\begin{equation}
  \delta {\cal R}_D (\Phi) \sim 
                \frac {A \varepsilon_{cl}^{-1}}
			      {
				   1 +
				  \displaystyle{
			        {\left( \frac{4 \pi} {\Theta_0 \varepsilon_{cl}} \frac {\Phi} {\Phi_0} \right)}^2
			      } 
				   }
 				 +
				 \sum_{n=1}^{\infty} 
	             \exp { 
				      \left[
                              -\left\{
							          \delta 
								       +
								     \frac {\beta} {2} 
								     {{\left( \frac{4 \pi} {\Theta_0} \frac {\Phi} {\Phi_0} \right)}^2} 
					            \right\} n
					  \right]
				     }
				 \cos{\left(4 \pi n \frac {\Phi} {\Phi_0}  \right)}, 
  \label{eqn:a13}
\end{equation}
where \( \Phi=B \Theta_0 / 2 \pi \).
The first term in eq. (\ref{eqn:a13}), $i.e$., 
\( 
A \varepsilon_{cl}^{-1} 
/
\left\{
1+
{\left( 2B / {\varepsilon}_{cl} \Phi_0 \right)}^2
\right\}
\), which contributes to 
negative magnetoresistance, agrees with Baranger, Jalabert and Stone's Lorentzian BWL correction
\begin{equation}
  \delta {\cal R}_D (B) = \frac{{\cal R}_{cl}} 
                               {
							     1 + 
								    \displaystyle{
								        \left( 
								              \frac{2B} {{\varepsilon}_{cl} \Phi_0} 
								        \right)^2
									}
								}
  \label{eqn:a19}
\end{equation}
for chaotic billiard ~\cite{rf:Baranger,rf:BJS}, where \( {\cal R}_{cl} \) 
is the wave-number-averaged classical reflection probability.

%
%
%
%===================================
\begin{center}
\begin{figure}[b]
%\vspace{-3.0cm}
\hspace{0.3cm}
\epsfxsize=8.5cm
\epsfbox{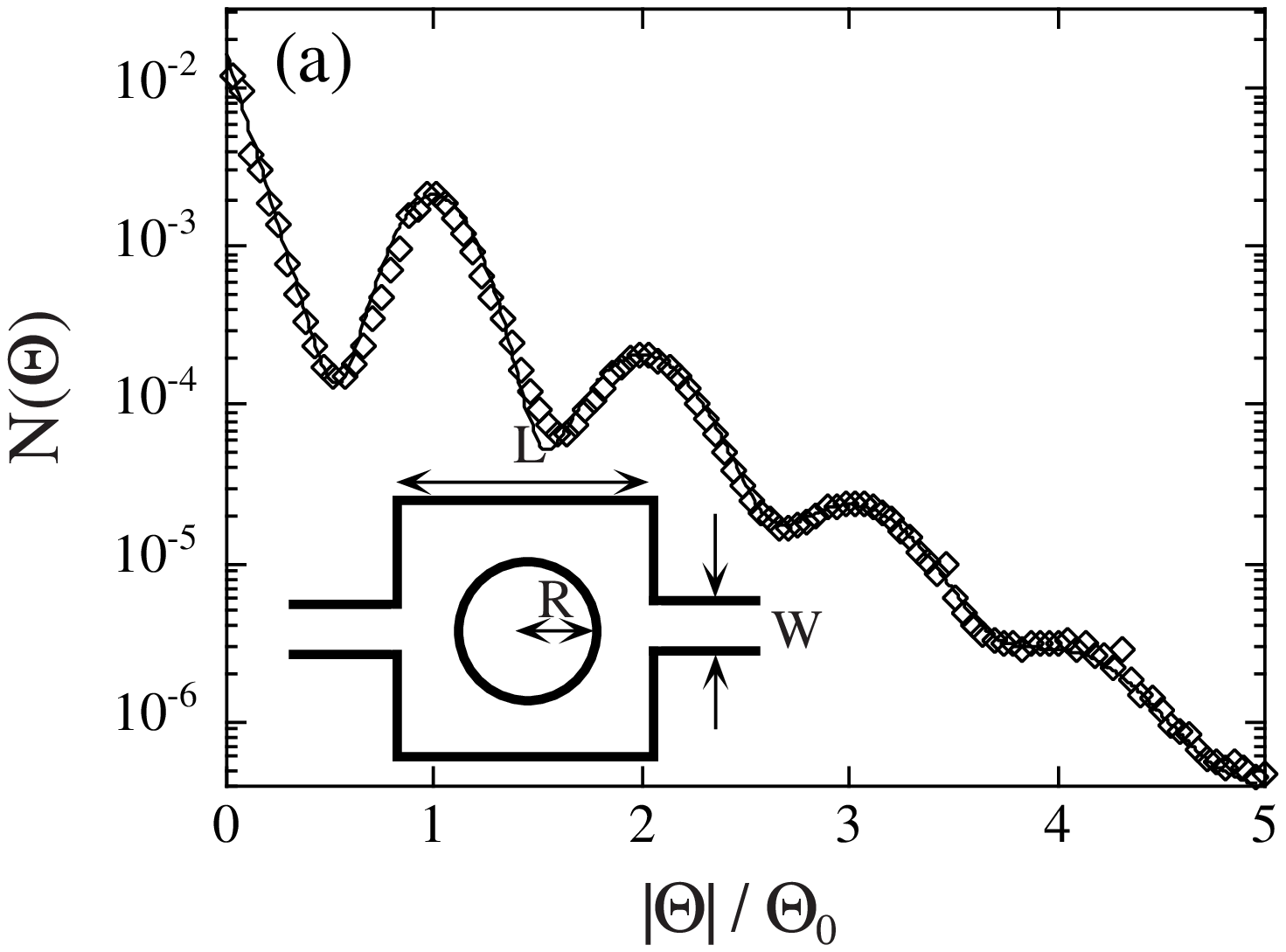}
\epsfxsize=8.5cm
\epsfbox{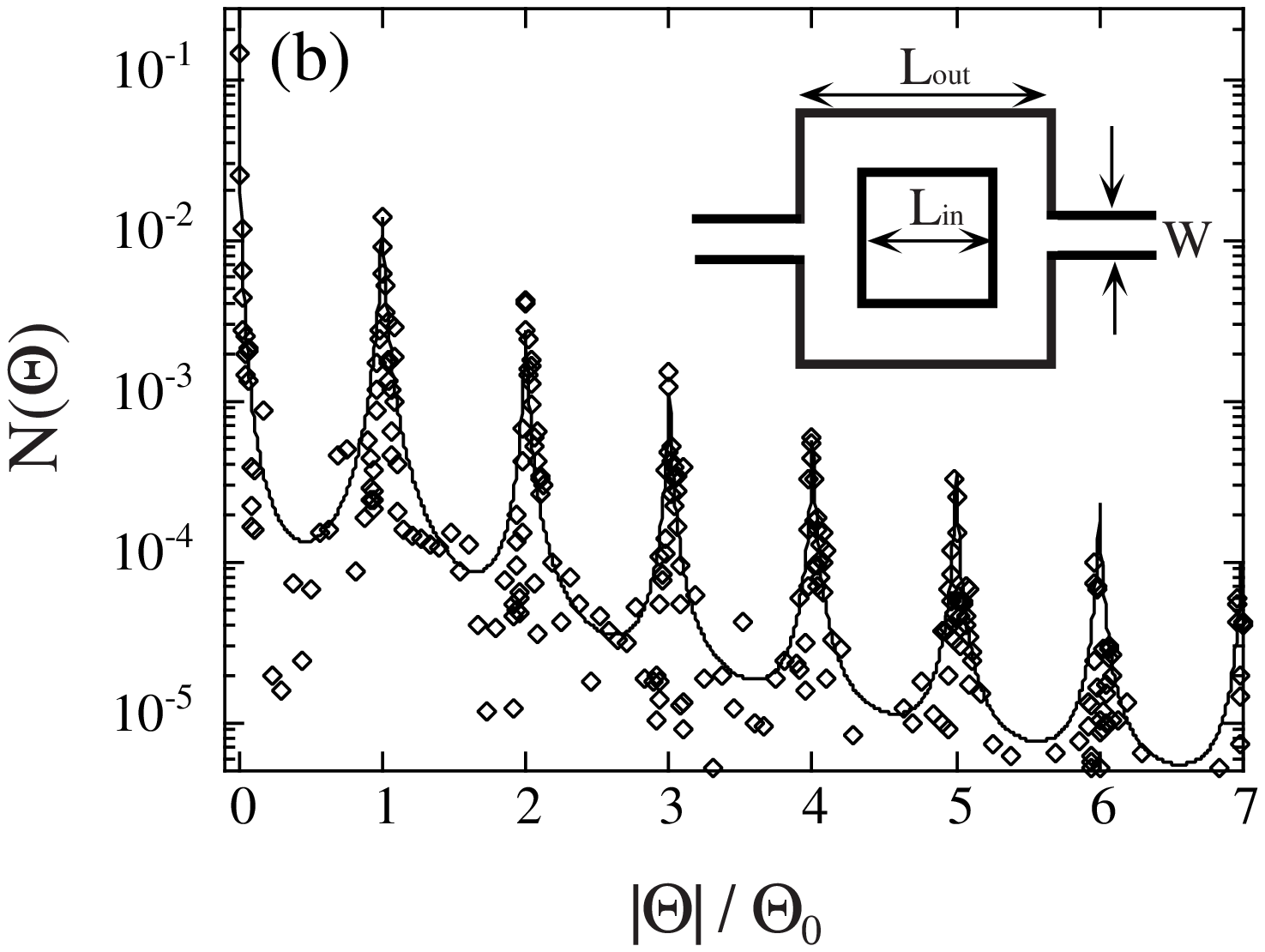}
\caption{Semi-logarithmic plot of the effective area distributions in scattering from the (a) Sinai (chaotic) billiard
(\(R/W=3\) and \(L/W=10\)) and (b) square AB (regular) billiard (\(L_{in}/W=6\) and \(L_{out}/W=10\)). 
The numerical simulation results (diamond) for (a) and (b) are 
well fitted by eqs. (13) and (16) (solid line), respectively.
The insets show the schematic views of two types of AB billiards.}
\end{figure}
\end{center}
%===================================
%
%
%

For the second term in eq. (\ref{eqn:a13}) (\( n>0 \) components), the oscillation amplitude decays exponentially 
with increasing $n$. Therefore, the main contribution to the conductance 
oscillation comes from the $n$=1 component, which oscillates with period \( h / 2e \). 
This behavior 
is consistent with our previous specific result [i.e., eq. (\ref{eqn:aas1})] for the chaotic AB billiard in which magnetic flux penetrates $only$ through 
the hollow ~\cite{rf:Kawabata}. 
In addition to this property, the oscillation amplitude 
damps exponentially with increasing
magnetic field.

 On the other hand, for regular cases (the square AB billiard)
the form of \( N(\Theta) \) 
has been estimated as
\begin{equation}
    N(\Theta) \sim n_0 
	            {\left(
				   \left| \Theta \right| + \Delta_1
				\right)}^{-z_2} 
				  +
	          \sum_{\stackrel{n=-\infty}{n \ne 0}}^{\infty} 
	            {\left(
				   \left| n \right| + n_1
				\right)}^{-z_1}  
	            {\left(
				   \left| \Theta -n \Theta_0 \right| + \Delta_2
				\right)}^{-z_2}  
  \label{eqn:a14}
\end{equation}
from the numerical simulation [see Fig. 1(b)]. 
In the calculation we injected \( 9 \times 10^8 \) particles into the billiard. 
In eq. (\ref{eqn:a14}) \( n_0, n_1 \), $z_1$, $z_2$, $\Delta_1$, and $\Delta_2$ 
are also fitting parameters 
and \( \Theta_0 / 2 \pi \) is approximately the average area between the outer and 
inner squares in this case.
This distribution leads to ~\cite{rf:comment2}  
\begin{equation}
  \delta {\cal R}_D (\Phi) \sim 
                n_0 A_1 (\Delta_1,\Phi) 
 				 +
				2 A_1 (\Delta_2,\Phi) \sum_{n=1}^{\infty} 
	            {\left( n + n_1 \right)}^{-z_1} 
				 \cos{\left(4 \pi n \frac {\Phi} {\Phi_0}  \right)}, 
  \label{eqn:a15}
\end{equation}
where
\begin{equation}  
  A_1(\Delta,\Phi) \equiv \int_{0}^{\infty} dx
                     (x+\Delta)^{-z_2} 
					 \cos{ \left(
					          \frac{4 \pi} {\Theta_0} \frac {\Phi} {\Phi_0} x 
					       \right) }.
  \label{eqn:a16}
\end{equation}
Since $A_1(\Delta_1,\Phi)$ is equal to the Fourier transform of a power-law function, one
can expect a cusplike BWL peak near zero magnetic field ~\cite{rf:Baranger,rf:Stone}. 
In contrast to chaotic cases, the oscillation amplitude decays algebraically for $n$.
Therefore, we can see that higher-harmonics components give a significant 
contribution to conductance oscillations. This is because the number of multiple-winding trajectories is much larger in regular
billiards than in chaotic billiards. 
\section{AAS OSCILLATION AND NEGATIVE MAGNETORESISTANCE}
 In this section we shall discuss in more detail the difference of \( \delta {\cal R}_D (\Phi) \) 
between chaotic and regular AB billiards.
In Fig. 2 we show \( \delta {\cal R}_D (\Phi) \) for Sinai (chaotic) 
and square AB (regular) billiards.
The values of the fitting parameters, determined by the classical simulation,
are substituted into eqs. (\ref{eqn:a13}) and (\ref{eqn:a15}).
In order to see the marked difference of the $\Phi$ dependence of \( \delta {\cal R}_D (\Phi) \), 
we shall investigate the $n=0$ term in eqs. (\ref{eqn:a13}) and (\ref{eqn:a15}),  
denoted as \( \delta {\cal R}_{NMR} (\Phi) \) and the \(n > 0\) term, 
denoted as \( \delta {\cal R}_{AAS} (\Phi) \), separately, i.e.,
%
%
%
%===================================
\begin{center}
\begin{figure}[b]
%\vspace{-3.0cm}
\hspace{0.3cm}
\epsfxsize=8.5cm
\epsfbox{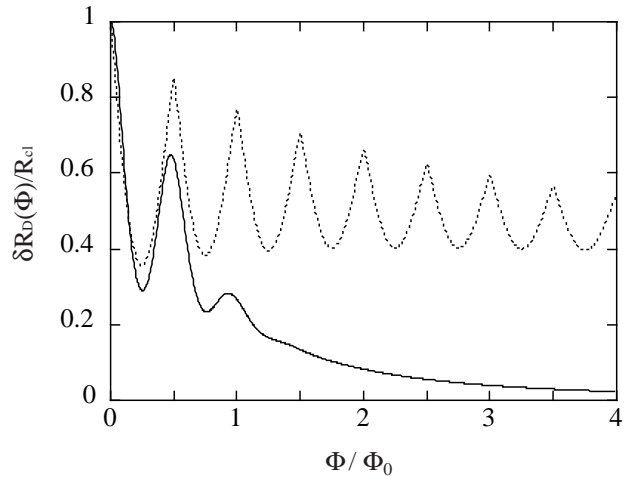}
\caption{Semi-classical BWL correction \(\delta {\cal R}_D \)
to the reflection coefficient
for chaotic AB (Sinai) billiard (solid) and regular square 
AB billiard (dotted) as a function of \(\Phi(=\Theta_0 B / 2 \pi)\).
\(\delta {\cal R}_D \) is normalized to the value at $\Phi=0$ 
, $i.e.,$ the classical reflection probability \({\cal R}_{cl} \).} 
\end{figure}
\end{center}
%===================================
%
%
%
% 
% 
\begin{equation}  
  \delta {\cal R}_D (\Phi)= \delta {\cal R}_{NMR}(\Phi) + \delta {\cal R}_{AAS}(\Phi).
  \label{eqn:a17}
\end{equation}
Figure 3(a) shows \( \delta {\cal R}_{NMR} (\Phi) \) which contributes to the negative 
magnetoresistance for two types of billiards. 
The shapes of \( \delta {\cal R}_{NMR} (\Phi) \) in the vicinity of zero magnetic field are quite different
 between chaotic and regular billiards, i.e., a quadratic curve versus a linear line 
 [see the inset in Fig. 3(a)].
For large $\Phi$ (but with the cyclotron radius sufficiently large compared to the system dimension),
\( \delta {\cal R}_{NMR} (\Phi) \) saturates in chaotic cases, but shows nosaturation in regular cases.
Similarly, the \( \delta {\cal R}_{AAS} (\Phi) \) corresponding to the $h/2e$ AAS-like oscillation part is indicated in Fig.3(b).
While the oscillation amplitude damps rapidly with increasing $\Phi$ for the chaotic AB billiard,
it damps gently for regular AB billiards.

Therefore, on the basis of the above results, the qualitative difference of \( \delta {\cal R}_D (\Phi) \) between 
chaotic and regular AB billiards is attributed to the
different classical distribution of the effective areas.
As the dimension of the center obstacle (e.g., $R$ for a Sinai billiard and $L_{in}$ for a square AB billiard) 
becomes zero, the oscillation structure of \( N(\Theta) \) is indistinct for two types of billiard, so that
the $h/2e$ conductance oscillation would disappear.

 To consolidate the above semiclassical prediction of the $h$/2$e$ oscillation, 
 we must compare 
 eqs. (\ref{eqn:a13}) and (\ref{eqn:a15}) with
 quantum-mechanical calculations (e.g., a recursive Green's function 
 method ~\cite{rf:Recursion}) 
 and also check the influence of the off-diagonal contribution 
 to \( \delta {\cal R} (\Phi) \) ~\cite{rf:Baranger,rf:BJS,rf:Takane}.
Moreover, it is desirable to confirm our prediction by having recourse to a
random matrix approach 
for systems with broken time reversal symmetries ~\cite{rf:RMT}.
Such investigations will be given elsewhere.
\section{Conclusion}
In summary, we have derived the semiclassical formula for \( \delta {\cal R}_D (\Phi) \) of single 
chaotic and regular AB billiards in which a weak $B$ is applied to the entire region.
We have shown that $h/2e$ oscillations and negative magnetoresistance would appear 
concurrently in 
\( \delta {\cal R}_D (\Phi) \):
as for {\em h/2e conductance oscillations}, we find the oscillation mainly with 
a fundamental period $h$/2$e$ and 
rapid damping of the amplitude with increasing $B$ for the chaotic billiard versus  
the large contribution of higher harmonic components and mild damping of the oscillation amplitude 
for a regular billiard
; As for {\em negative magnetoresistance},
the Lorentzian peak and saturation for the chaotic billiard versus a cusplike structure and no saturation for a 
regular billiard are reproduced.
Although Taylor $et$ $al.$ ~\cite{rf:Taylor} recently made an 
experimental study of the weak localization peak and self-similar structure of magnetoresistance
in a semiconductor Sinai billiard, no detailed experimental result of $h/2e$ 
conductance oscillations has yet been reported.
We hope that these characteristics of quantum chaos in the quantum magnetotransport will 
be experimentally observed in ballistic quantum dots forming AB geometry.
%
%
%
%
%===================================
\begin{center}
\begin{figure}[t]
%\vspace{-3.0cm}
\hspace{0.3cm}
\epsfxsize=8.5cm
\epsfbox{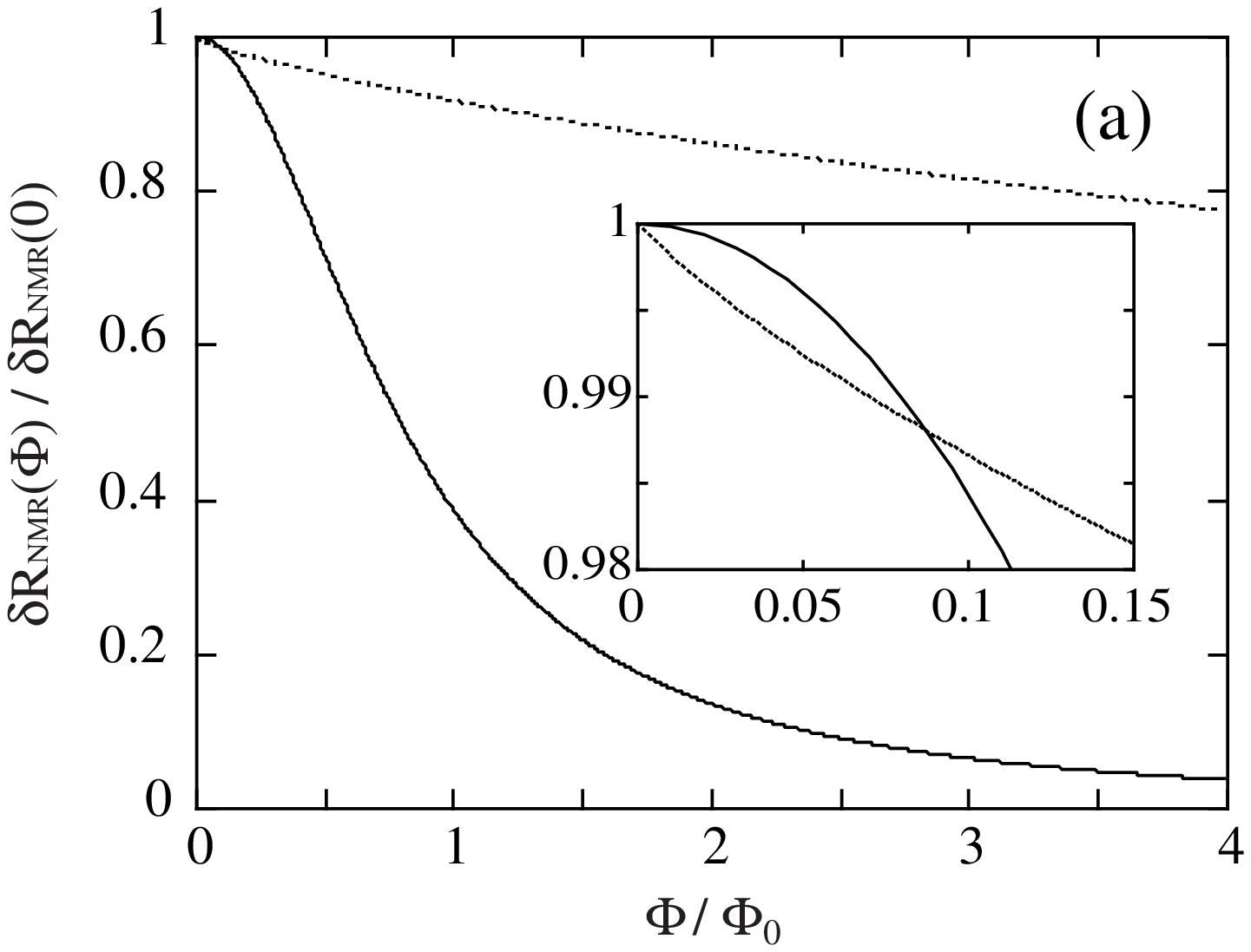}
\epsfxsize=8.5cm
\epsfbox{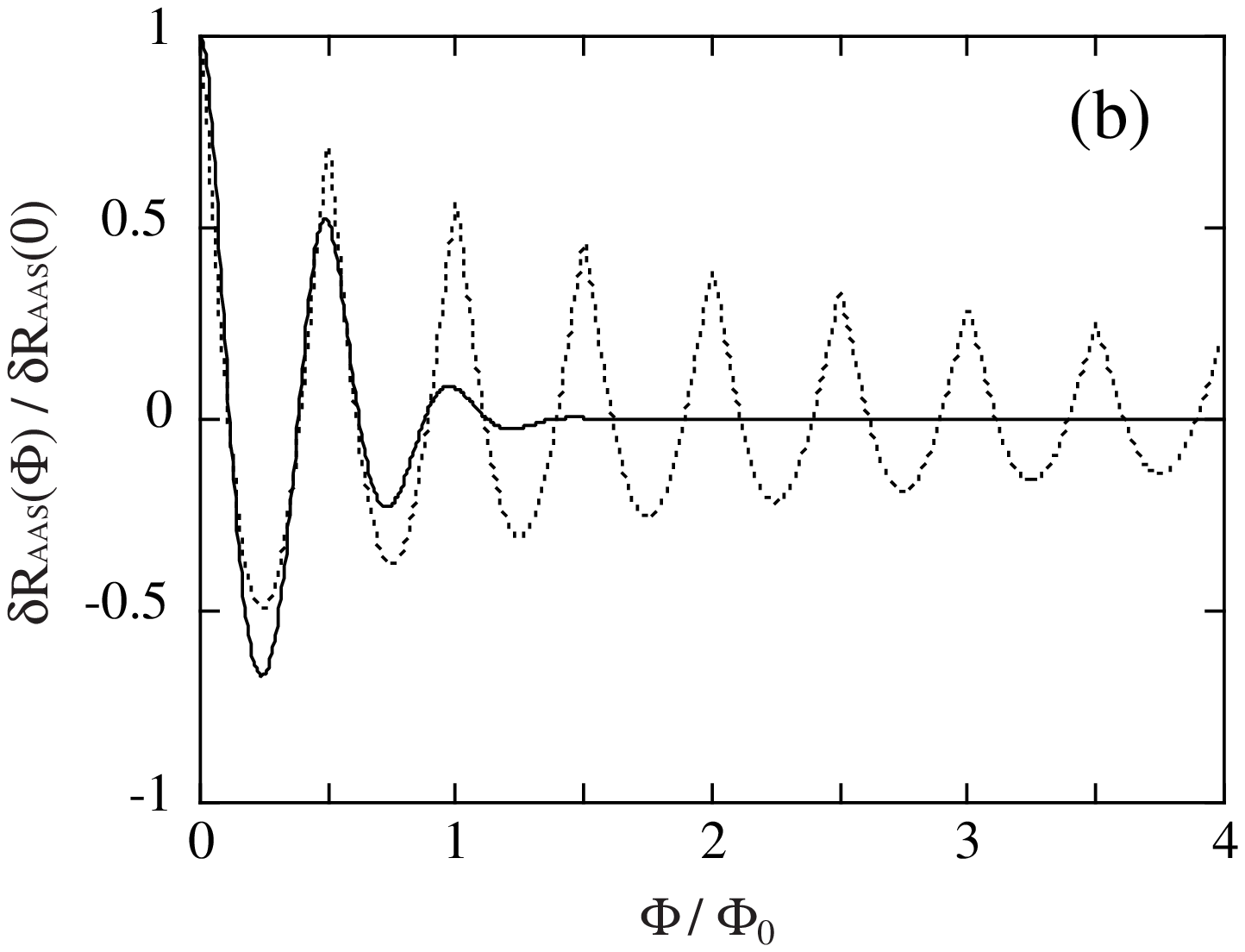}
\caption{Magnetic flux dependence of two different components of BWL corrections: (a) $\delta {\cal R}_{NMR}$
contributing 
to negative magnetoresistance and
(b) $\delta {\cal R}_{AAS}$ contributing to the $h/2e$ oscillation for
chaotic (solid) and regular (dotted) AB billiards.
\(\delta {\cal R}_{NMR} \) and \(\delta {\cal R}_{AAS} \) 
are normalized to the value at $\Phi=0$.
The inset in (a) shows $\delta {\cal R}_{NMR}$ in the vicinity of zero magnetic flux.}
\end{figure}
\end{center}
%===================================
%
%
%
\section{acknowledgements}
We would like to acknowledge 
Y. Takane, R.P. Taylor, Y. Ochiai, F. Nihey, W.A. Lin, and P. Gaspard
for valuable discussions and comments. 
Numerical calculations were performed on FACOM VPP500 in the Supercomputer 
Center, Institute for Solid State Physics, University of Tokyo.

\begin{references}
\bibitem{rf:comment2}Although the arguments leading to eq. (8) must be modified for regular 
systems [7], we can estimate the BWL correction term in such systems 
by simply substituting  eq. (16) into eq. (8).
\end{references}
\end{document}